# Boosting Simple Collaborative Filtering Models Using Ensemble Methods

Ariel Bar, Lior Rokach, Guy Shani, Bracha Shapira, Alon Schclar
Department of Information System Engineering,
Ben-Gurion University of the Negev Beer-Sheva 84105, Israel

**ABSTRACT**

In this paper we examine the effect of applying ensemble learning to the performance of collaborative filtering methods. We present several systematic approaches for generating an ensemble of collaborative filtering models based on a single collaborative filtering algorithm (single-model or homogeneous ensemble). We present an adaptation of several popular ensemble techniques in machine learning for the collaborative filtering domain, including bagging, boosting, fusion and randomness injection. We evaluate the proposed approach on several types of collaborative filtering base models: k-NN, matrix factorization and a neighborhood matrix factorization model. Empirical evaluation shows a prediction improvement compared to all base CF algorithms. In particular, we show that the performance of an ensemble of simple (weak) CF models such as k-NN is competitive compared with a single strong CF model (such as matrix factorization) while requiring an order of magnitude less computational cost.

## 1. INTRODUCTION

Collaborative Filtering is perhaps the most successful and popular method for providing predictions over user preferences, or recommending items. For example, in the recent Netflix competition, CF models were shown to provide the most accurate models. However, many of these methods require a very long training time in order to achieve high performance. Indeed, researchers suggest more and more complex models, with better accuracy, at the cost of higher computational effort.

Ensemble methods suggest that a combination of many simple identical models can achieve a performance of a complex model, at a lower training computation time. Various ensemble methods create a set of varying models using the same basic algorithm automatically, without forcing the user to explicitly learn a single set of model parameters that perform the best. The predictions of the resulting models are combined by, e.g., voting among all models. Indeed, ensemble methods have shown in many cases the ability to achieve accuracy competitive with complex models.

In this paper we investigate the applicability of a set of ensemble methods to a wide set of CF algorithms. We explain how to adapt CF algorithms to the ensemble framework in some cases, and how to use CF algorithms without any modifications in other cases. We run an extensive set of experiments, varying the parameters of the ensemble. We show that, as in other Machine Learning problems, ensemble methods over simple CF models achieve competitive performance with a single, more complex CF model at a lower cost.

## 2. BACKGROUND

We now review briefly basic needed concepts in collaborative filtering, ensemble methods, and some related work.

### 2.1 Collaborative Filtering

Collaborative Filtering (CF) [1] is perhaps the most popular and the most effective technique for building recommendation systems. This approach predicts the opinion that the active user will have on items or recommends the "best" items to the active user, by using a scheme based on the active user's previous likings and the opinions of other, like-minded, users.

The CF prediction problem is typically formulated as a triplet ($U$, $I$, $R$), where:

- $U$ is a set of $M$ users taking values form $\{u_1, u_2, \ldots, u_m\}$.
- $I$ is a set of $N$ items taking values from $\{i_1, i_2, \ldots, i_n\}$.
- $R$ - the ratings matrix, is a collection of historical rating records (each record contains a user id ($u \in U$), an item id ($i \in I$), and the rating that $u$ gave to $i$ –                .

A rating measures the preference by user $u$ to item $i$, where high values mean stronger preferences. One main challenge of CF algorithms is to give an accurate prediction, denoted by $\hat{r}_{u,i}$ to the unknown entries in the ratings matrix, which is typically very sparse. Popular examples of CF methods include k-NN models [1][2], Matrix Factorization models [3], and Naïve Bayes models [4].

### 2.2 Ensemble Methods

Ensemble is a machine learning approach that uses a *combination* of *identical* models in order to improve the results obtained by a single model. This approach has lately been receiving a substantial amount of research attention, due its effectiveness and simplicity. The ensemble model is constructed from a series of $K$ learned models (typically classifiers or predictors), $m_1, m_2, \ldots, m_k$, with the aim of creating an improved composite model $m^*$. Unlike hybridization methods [5] in recommender systems that *combine different types* of recommendation models (e.g. a CF model and a content based model), the base models which construct the ensemble are based on a single learning algorithm. For example, ensemble methods may invoke a *matrix factorization* algorithm several times, each time with different initial parameters \to receive a set of slightly different *matrix factorization* models, which are then combined to form the ensemble.

## 2.3 Related Work

Most improvements of collaborative filtering models either create more sophisticated models or add new enhancements to known ones. These methods include approaches such *matrix factorization* [3][6], enriching models with implicit data[7],enhanced *k-NN* models [8], applying new similarity measures [9], or applying *momentum techniques for gradient decent solvers* [6][10].

In [11] the data sparsity problem of the ratings' matrix was alleviated by imputing the matrix with artificial ratings, prior to building the CF model. Ten different machine learning models were evaluated for the data imputing task including decision tree, neural networks, vector machines and an ensemble classifier (a fusion of 7 of the previous 9 models). In two different experiments the ensemble approach provided lower *MAE*. Note that this ensemble approach is a sort of hybridization method.

The framework presented in [12] describes three matrix factorization techniques: *Regularized Matrix Factorization (RMF)*, *Maximum Margin Matrix Factorization (MMMF)* and *Non-negative Matrix Factorization (NMF)*. These models differed in the parameters and constraints that were used to define the matrix formation as an optimization problem. The best results (minimum *RMSE* measure) were achieved by an ensemble model which was constructed as a simple average of the three matrix factorization models.

Recommendations of several *k-NN* models are combined in [13]. The suggested model was a fusion between the *User-Based CF* approach and *Item-Based CF* approach. In addition the paper suggests lazy *Bagging* learning approach for computing the user-user, or item-item similarities. As reported, these manipulations improved the *MAE* in *k-NN* models.

In [14] a modified version of the *AdaBoost.RT* ensemble regressor a modification of the original *AdaBoost* [15] classification ensemble method designed for regression tasks) was shown to improve the *RMSE* measure of *a neighborhood matrix factorization* model. The authors demonstrate that adding more regressors to the ensemble reduces the RMSE (the best results were achieved with 10 models in the ensemble).

A heterogeneous ensemble model which blends five state-of-the-art CF methods was proposed in [16]. The hybrid model was superior to each of the base models. The parameters of the base methods were chosen manually.

The main contribution of this paper is a systematic framework for applying ensemble methods to CF methods. We employ *automatic* methods for generating an ensemble of collaborative filtering models based on a single collaborative filtering algorithm (homogeneous ensemble). We demonstrate the effectiveness of this framework by applying *several* ensemble methods to *various* base CF methods. In particular, we show that the performance of an ensemble of simple (weak) CF models such as k-NN is competitive compared with a single strong CF model (such as matrix factorization) while requiring an order of magnitude less computational cost.

## 3. ENSEMBLE FRAMEWORK

The proposed framework consists of two main components: (a) the ensemble method; and (b) the base CF algorithm. We investigate four common ensemble methods: Bagging, Boosting, Fusion (merging several models together, where each model uses the same base CF algorithm, but with different parameter values), and Randomness Injection. These methods were chosen due to their improved accuracy when applied to classification problems, and the diversity in their mechanisms. The first three approaches are general methods for constructing ensembles based on given any CF algorithm. The last one requires an adaptation of the CF algorithm that it uses.

The *Bagging* and *AdaBoost* ensembles require the base algorithm to handle datasets in which samples may appear several times, or datasets where weights are assigned to the samples (equivalent conditions). Most of the base CF algorithms assume that each rating appears only once, and that all ratings have the same weight. In order to enable application of *Bagging* and *Boosting*, we modify the base CF algorithms to handle recurring and weighted samples.

We evaluate five different base (modified) CF algorithms: *k*-NN User-User Similarity, *k*-NN Item-Item Similarity, Matrix Factorization (three variants of this algorithm) and Factorized Neighborhood. The first three algorithms are simpler, having a relatively low accuracy and rapid training, while the last two are more complex, having better performance and higher training cost.

## 4. ENSEMBLE METHODS FOR CF

We now provide a review of the ensemble methods which we use to demonstrate the proposed framework.

### 4.1 Bagging

The Bagging approach (Fig.1) [17] generates $k$ different bootstrap samples (with replacement) of the original dataset where each sample is used to construct a different CF prediction model. Each bootstrap sample (line 2) is in the size of the original rating data set, so some ratings may appear more than once, while others may not appear at all. The base prediction algorithm is applied to each bootstrap sample (line 3) producing $k$ different prediction models. The ensemble model is a simple average over all the base ones (line 5). This algorithm may work with every base CF prediction algorithm that can handle ratings with weights.

### 4.2 Boosting

AdaBoost [15] is perhaps the most popular boosting algorithms in machine learning. In this approach, weights are assigned to each rating tuple (initially, equal weights are given to all the examples). Next, an iterative process constructs a series of $K$ models. After model $M_i$ is learned, the weights are updated to allow the subsequent model, $M_{i+1}$, focus on the tuples that were poorly predicted by $M_i$. The ensemble model combines the predictions of each individual model via a weighted average, where the weight of each model is a function of its accuracy.

In this work we use a modified version of the *AdaBoost.RT* [18] algorithm. Specifically, we apply an absolute error function, rather than the traditional relative one. The learning algorithm receives four parameters: the first three are the original dataset, the base CF algorithm and the ensemble size - as in the Bagging approach. The fourth parameter $\delta$ – is a threshold value between 0 and the rating score range of the recommendation system (used as the demarcation criteria). The algorithm iteratively constructs the base models in the ensemble. In each iteration we use a different ratings' distribution – denoted by $D_t$ where $D_t(r_{ui})$ is the weight of the rating $r_{ui}$. Initially, the algorithm assigns the same weight to all ratings (lines 1-2).

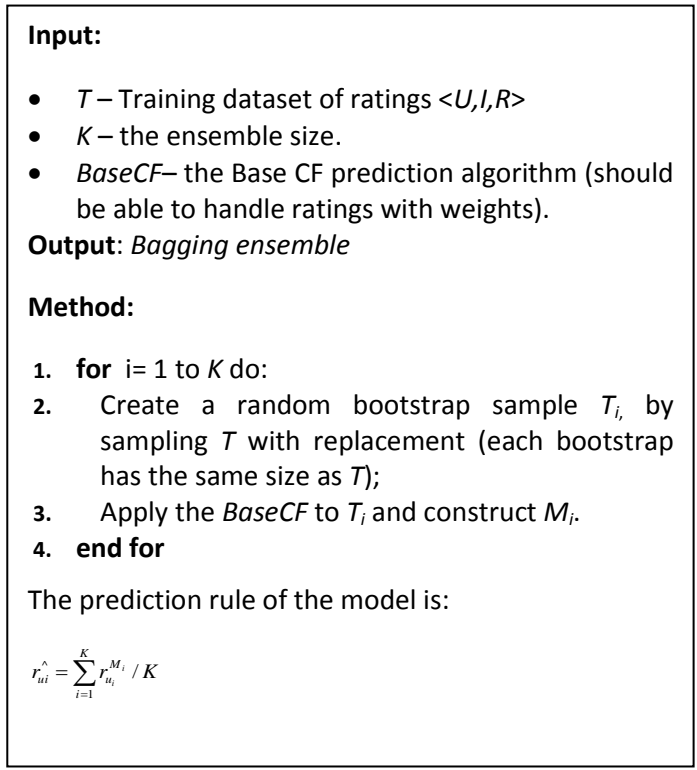

**Input:**

- *T* – Training dataset of ratings <U,I,R>
- *K* – the ensemble size.
- *BaseCF* – the Base CF prediction algorithm (should be able to handle ratings with weights).

**Output**: *Bagging ensemble*

**Method:**

1. **for** i= 1 to *K* do:
2. Create a random bootstrap sample $T_i$, by sampling *T* with replacement (each bootstrap has the same size as *T*);
3. Apply the *BaseCF* to $T_i$ and construct $M_i$.
4. **end for**

The prediction rule of the model is:

$$\hat{r_{ui}} = \sum_{i=1}^{K} r_{u_i}^{M_i} / K$$

**Figure 1: Bagging algorithm for CF**

The algorithm performs *K* iterations: First, the base model of the current iteration is constructed by applying *BaseCF* to the training set, with the current weight distribution (line 4). Second, the constructed model is evaluated by computing the absolute error (*AE*) of each rating in the dataset (lines 5-6) For example if the model predicts a 4.4 rating on a rating that it is actually 4, then the *AE* measure will be (4.4 – 4)= 0.4. Using the *AE* differs from the original algorithm, which applied a relative error function. Third, we calculate the total error rate $\varepsilon_t$ of the current model (line 7) - the summation of all the ratings' weights, which the model predicted incorrectly, i.e. their *AE* measure was above the threshold *δ*. Forth, we compute $\beta_t$ (the factor that is used to update the weight distribution) as the power *n* of $\varepsilon_t$ (line 8), where higher values of *n* indicates higher impact of $\varepsilon_t$ on the ensemble. In this work we use *n=1*. Finally, in line 9 we update the distribution for the next iteration (increase the weights of the ratings which were predicted incorrectly).

The prediction rule of the ensemble is a weighted average over all the base models in the ensemble. The weight of each model is based on the value of its $\beta_t$, where large values mean less weight contributed to that model in the ensemble.

As suggested in the original algorithm, we initialize *δ* to be the *AE* of the original dataset.

### 4.3 Fusion

A straightforward way to construct an ensemble is to take a specific prediction algorithm, and use it several times on the same dataset, but each time with different initial parameters [19]. This process constructs different models, which can later be combined together by e.g. averaging. For example, different matrix factorization models may be built using different sizes of latent factors. A simple fusion of these models can be calculated as the average of their outputs. Figure 3 summarizes the Fusion approach.

---

**Input:**

- *T* – Training dataset of ratings <U,I,R>
- *K* – the ensemble size.
- *BaseCF*– the Base CF prediction algorithm (should be able to handle ratings with weights).
- *δ* – Threshold (0 <*δ*<*the rating score range*) for demarcating correct and incorrect predictions

**Output**: *AdaBoost.RT model*

**Method:**

1. Assign iteration number *t=1*
2. Assign initial distribution for each tuple in *R*:
   $D_t(r_{ui}) = 1/|R|$
3. **while**  t ≤ *K*  Do
4.      Apply *BaseCF* to *T* with distribution $D_t$, and construct the model $M_t$.
5.      **for each** rating $r_{ui} \in$ R
6.           calculate $AE_t(r_{ui}) = |r_{ui}^{\wedge Mt} - r_{ui}|$
7. calculate error rate of iteration *t*:
   $$\varepsilon_t = \sum_{ui: AE_t(r_{ui}) > \delta} D_t(r_{ui})$$
8. Set $\beta_t = \varepsilon_t^n$
9. Update distribution $D_{t+1}$ as:
   $$D_{t+1}(r_{ui}) = \frac{D_t(r_{ui})}{Z_t} \times \begin{cases} \beta_t & \text{if } AE_t(r_{ui}) \leq \delta \\ 1 & \text{otherwise} \end{cases}$$
   //$Z_t$ is a normalization factor for keeping the weights as a distribution
10. set *t = t+1*
11. **end while**

The prediction rule of the model is:
$$\hat{r_{ui}} = \sum_{i=1}^{K} \log(\frac{1}{\beta_i}) \cdot r_{ui}^{\wedge Mi} / \sum_{i=1}^{K} \log(\frac{1}{\beta_i})$$

---

**Figure 2: AdaBoost.RT algorithm for CF**

In this work the fusion manipulation was as follows:

- For *k*-NN, we applied the following fusion schemes:

1. *k-NN fusion by similarity metric*–we combined the predictions of a *k-NN* model based on the *Pearson* correlation similarity and a *k-NN* model based on the *Cosine* similarity. The same values were used for the other parameters of both models (*k*-NN size, aggregation function, User/Item perspective).

2. *k-NN Fusion by CF perspective* – in this schema we combined the predictions of the *User-User k-NN* model, and the *Item-Item k-NN* model. The same values were used for the other parameters of both algorithms (k-NN size, aggregation function, similarity metric).

3. *k-NN Fusion by C F perspective & similarity metric*– this is an ensemble of four *k*-NN models: *k-NN-User-User-Pearson + k-NN-User-User-Cosine + k-NN-Item-Item-Pearson + k-NN-Item-Item-Cosine*. Basically this is a combination of the two previous fusion schemes.

For matrix factorization, we applied fusion to models which were constructed using different vector sizes of the latent factors.

**Input:**

- *T* – Training Dataset of ratings given as *<U,I,R>*
- *K* – the ensemble size.
- *BaseCF*– the Base CF prediction algorithm.
- *ParametersSet* – a set of different parameters that can be applied to *BaseCF*.

**Output**: *Fusion model*

**Method:**

1. **for** i=1 to *K* do
2. Apply *BaseCF* to *T* with the parameters in *Parameters_Set* [i], and construct model $M_t$.
3. **end for**

The final prediction rule is:

$$\hat{r_{ui}} = \sum_{i=1}^{K} r_{ui}^{\hat{}Mi} / K$$

**Figure 3: Fusion schema for CF**

**4.4 Randomness Injection**

All ensemble methods described so far in this section are generic in the sense that they are not limited to a specific CF prediction algorithm. Thus one of their parameters is *BaseCF* – the base CF prediction algorithm, which is used to construct the base models in the ensemble.

A different approach to create an ensemble is to take a base algorithm and modify it such that it will create various sub models and combine their results. A popular way to achieve this is by introducing randomness to the basic learning schema. By doing so, it is possible to run the algorithm several times, and receive a different model each time. These models can then be joined to provide a combined prediction. In this work the randomness was injected as follows:

- Random $k$-NN - Instead of selecting the top $k$ nearest neighbors (users or items) for the prediction rule, we randomly select any $k$ users/items from the top $2*k$ nearest neighbors. We can repeat this process $K$ times (the ensemble size) to get $K$ different predictions, and then use a simple average on them for the final one.
- Random Matrix Factorizations/Weighted Factorized Neighborhood–The MF algorithms are naturally randomized, since in the initialization process of the learning phase, we assign small random numbers to the latent factors. If we simply repeat this process $K$ times (the ensemble size), each time with random initial values, we will receive $K$ different MF models. These models can then be combined into an ensemble by a simple average.

## 5. MODIFIED CF ALGORITHMS

Some ensemble methods require that the base prediction algorithm can handle datasets with reoccurring or weighted samples. Accordingly, we had to modify CF algorithms which assume that each rating appears once, and that all ratings weights are equal. The first step in our modification was to update the original CF prediction problem from Section 2.1, by adding a new element $W$ to the problem formalization. $W$ is a vector of weights whose size is equal to the number of ratings where $w_{u,i}$ $\in W$ indicates the relative distribution (weight) of rating $r_{u,i}$ over $R$. In the following section we describe the modifications made to the original CF algorithms, in order to take into consideration the new weights vector. It is important to notice that when all weights are equal, all modified algorithms coincide with the original ones.

### 5.1 Modified k-NN Algorithms

The prediction rule of the modified $k$-NN User-User and the $k$-NN Item-Item algorithms [1][2] is similar to the original rule: the algorithm receives as input the neighborhood size, the similarity measure (user-user or item-item) and the final prediction aggregation function. Based on the similarity function the *k-nearest neighbors* are found the final prediction is derived. Our modifications focus on the similarity measures and aggregation functions.

For the user-user *k*-NN prediction, we suggest using the modified *Pearson correlation coefficient*, and the modified *cosine-based* user-user similarity measures as described in Eqs.1 and 2 respectively, where $S(u,v)$ is the set of items that both users $u$ and $v$ rated, $r_u$ is the weighted average rating of user $u$, and $w_{uvj}$ is the maximum between $w_{uj}$ and $w_{vj}$.

$$pearson(u,v) = \frac{\sum_{j \in S(u,v)} w_{uvj}(r_{u,j} - r_u)(r_{v,j} - r_v)}{\sqrt{\sum_{j \in S(u,v)} w_{uvj}(r_{u,j} - r_u)^2 \sum_{j \in S(u,v)} w_{uvj}(r_{v,j} - r_v)^2}} \quad (1)$$

$$\cos ine(u,v) = \cos(\vec{u},\vec{v}) = \frac{\vec{u} \bullet \vec{v}}{\|\vec{u}\|_2 \times \|\vec{v}\|_2} = \frac{\sum_{j \in S(u,v)} w_{uvj} r_{u,j} r_{v,j}}{\sqrt{\sum_{j \in S(u,v)} w_{uvj} r_{u,j}^2} \sqrt{\sum_{j \in S(u,v)} w_{uvj} r_{v,j}^2}} \quad (2)$$

This adjustment gives more emphasis to ratings with higher weights when calculating user-user similarities. For example, suppose we compare the similarity between "user 1" and "user 2", when the common item is "item 3", when $w_{13}$=2 and $w_{23}$=4 (meaning that the rating that "user 1" gave to "item 3" appears 2 times in the dataset and the rating that "user 2" gave to "item 3" appears 4 times), then the value of $w_{123}$ will be 4.

We aggregate the neighbors' votes using the modified *adjusted weighted average*, which takes into consideration the relative weights of each rating in the dataset as presented in equation 3. Note that $r_u$ denotes the weighted average of user $u$ ratings and        is the set of $k$ nearest neighbors of the user $u$.

$$\hat{r_{u,i}} = r_u + \frac{1}{\sum_{v \in N_k(u)} |sim(u,v)|} \sum_{v \in N_k(u)} sim(u,v) \cdot (r_{v,i} - r_v) \quad (3)$$

For the item-item *k*-NN algorithm we used the same modified functions by exchanging the indices.

**5.2  Modified MF Algorithms**

In this work we modified the ISMF, RISMF and BRISMF algorithms from [6] to handle weighted datasets. The modified algorithms apply a new optimization problem as presented in Eq.4. The new optimization problem seeks the optimal P* and Q* matrices (users matrix and items matrix, respectively) that minimize the SSE (Sum of Square Errors) with respect to the weight of each sample, while trying to avoid overfitting by using λ as a small non-negative realization parameter.

$$e_{ui} = (\hat{r}_{ui} - r_{ui})$$

$$e'_{ui} = w_{ui} \frac{e_{ui}^2 + \lambda \cdot p_u \cdot p_u^T + \lambda \cdot q_i \cdot q_i^T}{2} \quad (4)$$

$$SSE' = \sum_{(u,i) \in T} e'_{ui}$$

$$(P^*, Q^*) = \arg\min_{(P,Q)} SSE'$$

Where and are the ratings of user *u* and item *i*, respectively.

To solve the new optimization problem we apply the same gradient descent method that was used in the original algorithms (including the same *η* as a small non-negative step parameter). The only modification is in applying the new gradient steps according to the modified optimization problem as follows:

$$p'_{uk} = p_{uk} + \eta \cdot w_{ui} \cdot (e_{ui} \cdot q_{ki} - \lambda \cdot p_{uk})$$

$$q'_{ki} = q_{ki} + \eta \cdot w_{ui} \cdot (e_{ui} \cdot p_{uk} - \lambda \cdot q_{ki})$$

### 5.3 Modified Factorized Neighborhood Algorithm

We modify the Factorized Neighborhood Model (*FNM*) [8] to handle weighted datasets. We chose this model due to the high accuracy of its predictions. It is also interesting that FNM combines the latent factors paradigm with a *k-NN* formation.

The first modification to the algorithm is in updating the prediction rule:

$$\hat{r}_{ui} = \mu + b_u + b_i + q_i^T (|R(u)|^{-\alpha} \sum_{j \in R(u)} w_{uj}(r_{uj} - b_{uj})x_j + \quad (5)$$
$$|N(u)|^{-\alpha} \sum_{j \in N(u)} w_{uj} y_j)$$

The new prediction rule is consistent with the original while taking into account the rating's weights in all summations. The second modification is in the optimization problem:

$$\min_{q^*,x^*,y^*,b^*} \sum_{(u,i)\in K} w_{ui}((r_{ui}-\mu-b_u-b_i \qquad (6)$$
$$-q_i^T(|R(u)|^{-\alpha}\sum_{j\in R(u)}w_{uj}(r_{uj}-b_{uj})x_j+|N(u)|^{-\alpha}\sum_{j\in N(u)}w_{uj}y_j))^2$$
$$+\lambda(b_u^2+b_i^2+\|q_i\|^2+\sum_{j\in R(u)}\|x_i\|^2+\sum_{j\in N(u)}\|y_i\|^2))$$

The differences are in:

- Adding $w_{ui}$ to the optimization problem to minimize the overall error of the model with respect to the weights of the ratings.

- Initializing the baseline estimators with weighted, instead of non-weighted, averages.

As in the MF algorithms from Section 5.2, the original FNM learning algorithm solves the optimization problem by applying gradient descent. We apply the same method to solve the new optimization problem, so the third and last modification is in applying the new gradient steps:

$$q_i = q_i + \eta \cdot w_{ui}(e'_{ui} \cdot p_u - \lambda \cdot q_i)$$

$$b_u = b_u + \eta \cdot w_{ui}(e'_{ui} - \lambda \cdot b_u)$$

$$b_i = b_i + \eta \cdot w_{ui}(e'_{ui} - \lambda \cdot b_i)$$

$$x_i = x_i + \eta \cdot w_{ui} \cdot (|R(u)|^{-\alpha} \cdot (r_{ui}-b_{ui}) \cdot sum\_error - \lambda \cdot x_i)$$

$$y_i = y_i + \eta \cdot w_{ui} \cdot (|N(u)|^{-\alpha} \cdot sum\_error - \lambda \cdot y_i)$$

## 6. EVALUATION

### 6.1 Experimental Setup

The evaluation of the algorithms described in sections 4 and 5 was mainly based on the 100K *MovieLens* dataset. We used RMSE for measuring accuracy. We use 5 different random 80:20 splits over the original dataset. All algorithms ran on the same splits. We compare the following configurations:

- The *k-NN User-User Similarity* algorithm from subsection 5.1 (*k-NN-User*) was evaluated by applying 2 different modified similarity measures (*Pearson* and *Cosine*), and 3 k-NN sizes (5, 10, 20) producing a total of 6 different configurations.

- The *k-NN Item-Item Similarity* algorithm from subsection 5.1 (*k-NN-User*) was evaluated over the same configurations as the *k-NN-User* model.

- The three *matrix factorization* algorithms from Section 5.2 (*ISMF, RISMF, BRISMF*) were each evaluated using different sizes of latent factors vectors (3, 4, 5, 10, 20, 30, 40, 50). All latent factor vectors were initialized with small random numbers from the interval [-0.01, 0.01], the regulation parameter $\lambda$ was set to 0.01, and the step parameter $\eta$ was set to 0.01. These parameters' values are consistent with the original models.

- The *Factorized Neighborhood Mode l*algorithm from susection 5.3 (*FNM*) was evaluated for the latent factor vectors sizes: 3, 4, 5, 10, 20 and 30. All latent factors vectors were initialized with small random numbers from the interval [-0.01, 0.01], the regulator parameter $\lambda$ was set to 0.04, the step parameter $\eta$ was set to 0.002, the normalization factor α was set to ½ and the number of iterations was set to 20, as in the original model.

For each configuration we evaluated its original RMSE without any ensemble enhancement.We use these results as a baseline. We apply all ensemble methods from Section 4 to each configuration with different ensemble sizes and compared the results to the baseline:

- *Bagging* – for the k-NN algorithms, we apply ensemble sizes of 5, 10 and 20, and for the MF algorithms we applied ensemble sizes of 5, 10, 20, 30, 40 and 50.
- *AdaBoost.RT* – for all the algorithms, we apply ensemble sizes ranging from 1 to 10.
- *Fusion*:
    1. For the *k*-NN algorithms we apply the three ensemble schemas from subsection 4.3.
    2. For the MF algorithms we apply ensemble sizes of 5 and 10.
- *Randomness Injection* – we apply different ensemble sizes ranging from 1 to 10 for all algorithms.

### 6.2 Results

We now report various results and insights from this experiment: accuracy results over all ensemble configurations, the effect of ensemble size on scalability, and more.

#### 6.2.1 Accuracy Results

Due to space restrictions, we are unable to report all possible RMSE results. We therefore limit Table 1 to the best configuration of each method. For example, from all k-NN User-User ensemble configurations using Bagging, the ensemble over *k=20* produced the best results and is hence reported in the table.

We organized the base CF model according to their relative "strength", where simple/less accurate models appear on the left, and more advanced/complex/accurate appear on the right. The final row of the table indicates the improvement percentages of the best ensemble model compared to the baseline model.

**Table 1**: Accuracy Results (RMSE). The RMSE is given along with the size (in parentheses) of the ensemble which achieved it.

| CF Algorithm \ Ensemble | k-NN-User | k-NN-Item | ISMF | RISMF | BRISMF | FNM |
|---|---|---|---|---|---|---|
| *Baseline* | 0.9535 | 0.9526 | 0.9434 | 0.9407 | 0.9268 | 0.9231 |
| *Bagging* | 0.9495* (20) | 0.9464* (20) | **0.9173*S (50)** | **0.9152*S (50)** | **0.9170*S (50)** | *0.9333* |
| *AdaBoost.RT* | 0.9410*S (10) | 0.9459* (10) | 0.9397* (10) | *0.9415* | *0.9332* | *0.9367* |
| *Fusion* | **0.9383*S (4)** | **0.9383*S (4)** | 0.9411* (10) | 0.9383* (10) | 0.9241* (10) | 0.9158* (10) |
| *Random* | 0.9462* (10) | 0.9437* (10) | 0.9407* (10) | 0.9381*(10) | 0.9237*(10) | **0.9153* (10)** |
| *Improvement* | 1.57% | 1.47% | 2.76% | 2.66% | 0.97% | 0.87% |

We use the following notations in the table: ensemble enhancements which improved with statistical significance the RMSE measure over the baseline accuracy are denoted with "*". The best model in each column is displayed in bold-face font. Ensemble models of relatively weak algorithms which improve the RMSE to a level of more advanced models are denoted with "S". The ensemble sizes indicated inside the parentheses of each ensemble configuration.

We check for statistical significance using *One-Way ANOVA* with repeated measures (applying the *Greenhouse-Geisser test*) with confidence level α = 0.05, followed by a simple *paired t-test*, with confidence levelα= 0.05. Our results indicate the following:

1. We were able to significantly improve the baseline results of every base CF model type in our work, by at least two different ensemble approaches.
2. The improvement level was between 0.87% and 2.76%. These improvements may seem modest, but lowering the RMSE is a difficult problem, and every reduction in RMSE is difficult to achieve.
3. The improvement level depends on the base CF models - more complex models are more difficult to improve. This agrees with the idea that ensemble should be applied to boost the performance of weak CF models, not to improve complex models.
4. The *Fusion* and *Random Injection* ensemble methods were able to improve the accuracy of all base CF models.

5. *Bagging* failed to improve *FNM*, and *AdaBoost* failed to improve *RINMF*, *BRISMF* and *FNM*, however, these ensemble approaches may achieve better results than *Fusion* or *Randomness Injection* when applied to other base CF algorithms.

6. In the spirit of the "No Free Lunch" theorem, none of the evaluated ensemble method was optimal for all given scenarios. Consequently, one should look for the (base model, ensemble) pair that achieves the best results for the dataset at hand.

*6.2.2 The Effect of the Ensemble Size*

Table 1 shows that if the ensemble method (any method of the evaluated four) improves the accuracy of the basic model, then the ensemble model that achieved the best result is the one with the highest number of members. In these cases the overall RMSE of the ensemble decreased, as the number of members increased. Consequently, the strategy in this case is to use as many ensemble members as possible provided that the improvement is significant, and feasible with the amount of computation resources. Figure 4 demonstrates this idea by using Randomness injection on FNM.

Adding more members to the ensemble may be practical, as the complexity of all the ensemble methods grows linearly in the number of ensemble members. This is especially true in the case of parallel computing, where different ensemble methods can be trained on different cores or machines (all except for AdaBoost) Figure 5 shows this by using randomness injection on FNM: the learning and prediction times of the ensemble model grows linearly with the ensemble size.

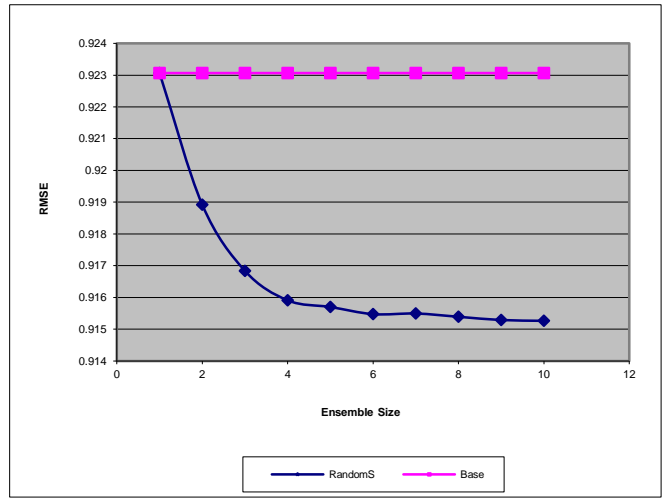

**Figure 4: FNM vs Random FNM (RMSE results)**

*6.2.3 Computational Cost and Accuracy Tradeoff*

As described in sub-section 6.2.1 in several scenarios an ensemble of relatively *weak* models achieved better accuracy than a single *stronger* model. Figure 6 present the RMSE obtained by various methods as function of the computation cost (training time – presented in log scale). The graph in Figure 6 shows the following results:

- An ensemble of the k-NN-User method achieves competitive performance with two MF methods (ISMF and RISMF) at an order of magnitude less computational cost (4 seconds instead of 24-26).
- An ensemble of MF methods (ISMF and RSIMF) achieves a competitive performance with a BRISMF method at a much lower computational cost (170 seconds instead of 490).

*6.2.4 Additional Accuracy Results*

We now present results for the larger MovieLens dataset (denoted as "MLB"), which contains 1 million ratings. Due to time limitations it was not feasible to test all methods. Therefore, for MLB, for each of base CF models, we evaluated only the two ensemble models which produced the best results on the MovieLens 100K dataset. In addition, for each base CF model we evaluated the ensemble enhancements only with respect to the best base configuration. Table 2 summarizes the RMSE results of the experiments in this section. For each ensemble model we specify the ensemble size in parentheses. Ensemble models which improved the base model are denoted with [I], ensemble models which improved the *RMSE* of a "weaker" base CF model to the level of "stronger" base CF models are denoted with [S]. The best model for each base CF model/dataset is bolded. In these experiments we applied the same configurations as described in sub-section 6.1 except for the maximum ensemble size that was set to 30 in the Bagging experiments with MF algorithms. The accuracy results in this experiment are consistent with the ones in previous sections, except for fusion of *k*-NN which was unable to improve the overall accuracy of the model to the level of a single ISMF or RISMF one.

## 7. CONCLUSIONS

In this work we presented a novel systematic framework for applying ensemble methods to collaborative filtering models. Our framework used four popular ensemble techniques (*Bagging, Boosting, Fusion and Randomness Injection*) which were adapted to solve the collaborative filtering based rating prediction task. Typical collaborative filtering algorithms neither handle datasets with reoccurring samples, nor weighted samples. We thus modify the original base collaborative filtering algorithms to handle such settings.

Empirical evaluation shows an RMSE improvement by applying the suggested ensemble methods to the base CF algorithms. These improvements may increase the accuracy of relatively weak models to the level of more advanced ones. We found that in most cases it is preferable to add more base models to the ensemble, as we obtain a more accurate model compared to the combined model. Since all our ensemble methods have a linear running time and space complexity with respect to the ensemble size, it may be feasible to add more models to the ensemble as long as the improvement level is significant. These encouraging results indicate that ensemble methods can be used to enhance collaborative filtering algorithms. In the future we plan to evaluate our suggestions on other datasets and also on other recommendation tasks.

A key issue that needs further investigation is how to find a data-driven criterion for choosing the optimal (ensemble, base model) pair for a given dataset. Other issues that need to be addressed are: evaluation of additional collaborative filtering models; application of ensemble methods to other recommendation systems types (content-based, demographic-based and hybrid-based); and the application of other ensemble methods.

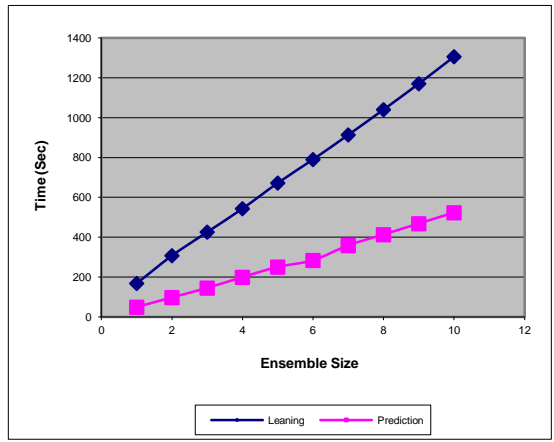

**Figure 5: Running time is linear in the size of the ensemble**

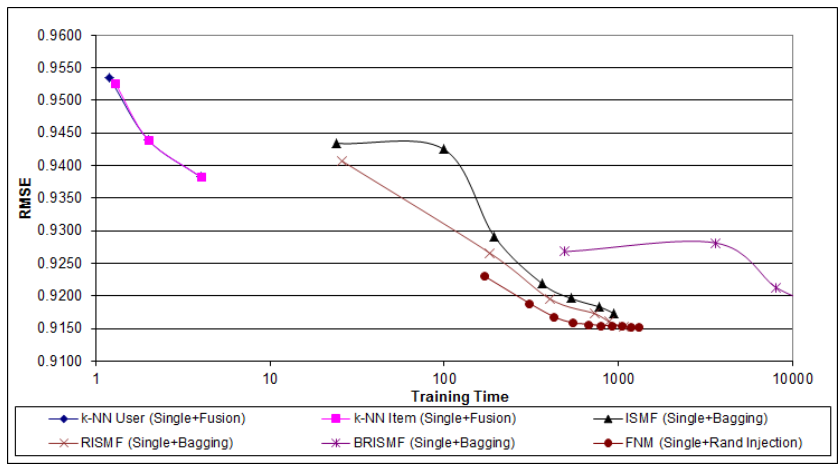

**Figure 6: Computational cost Vs. RMSE**

Table 2: MLB Accuracy Results (RMSE)

| BaseCF | Ensemble Model | RMSE-MLB (Ensemble Size) |
|---|---|---|
| *KNN-User* | *Base (None)* | 0.9302 |
| | *Fusion* | **0.8972 (4)** [I] |
| | *AdaBoost.RT* | 0.9116 (10) [I] |
| *KNN-Item* | *Base (None)* | 0.9029 |
| | *Fusion* | 0.8972 (4) [I] |
| | *Random* | **0.8954 (10)** [I] |
| *ISMF* | *Base (None)* | 0.8812 |
| | *Bagging* | **0.8523 (30)** [I S] |
| | *Random* | 0.8759 (10) [I] |
| *RISMF* | *Base (None)* | 0.8712 |
| | *Bagging* | **0.8480 (30)** [I S] |
| | *Random* | 0.8673 (10) [I] |
| *BRISMF* | *Base (None)* | 0.8620 |
| | *Bagging* | **0.8519 (30)** [I] |
| | *Random* | 0.8570 (10) [I] |
| *FNM* | *Base (None)* | 0.8654 |
| | *Fusion* | 0.8469 (10) [I] |
| | *Random* | **0.8465 (10)** [I] |

## 8. REFRENCES